\documentclass[prl,twocolumn,aps]{revtex4}
\usepackage{amsfonts}
\usepackage{amsmath}
\usepackage{graphicx}
\usepackage{xcolor}

\begin{document}

\title{Experimental Fock-State Superradiance}

\author{L. Ortiz-Guti{\'e}rrez$^{1}$, L. F. Mu{\~n}oz-Mart{\'i}nez$^{1}$, D. F. Barros$^{2}$, J. E. O. Morales$^1$, \\ R. S. N. Moreira$^1$, N. D. Alves$^1$, A. F. G. Tieco$^1$, P. L. Saldanha$^2$, and D. Felinto$^1$}

\affiliation{$^{1}$Departamento de F\'{\i}sica, Universidade Federal de Pernambuco, 50670-901 Recife-PE, Brazil \\ $^{2}$Departamento de F\'isica, Universidade Federal de Minas Gerais, 30161-970 Belo Horizonte-MG, Brazil}

\begin{abstract}
Superradiance in an ensemble of atoms leads to the collective enhancement of radiation in a particular mode shared by the atoms in their spontaneous decay from an excited state. The quantum aspects of this phenomenon are highlighted when such collective enhancement is observed in the emission of a single quantum of light. Here we report a further step in exploring experimentally the nonclassical features of superradiance by implementing the process not only with single excitations, but also in a two-excitations state. Particularly we measure and theoretically model the wave-packets corresponding to superradiance in both the single-photon and two-photons regimes. Such progress opens the way to the study and future control of the interaction of nonclassical light modes with collective quantum memories at higher photon numbers.
\end{abstract}

\maketitle

The full quantum mechanical treatment of spontaneous emission from an ensemble of atoms may lead to enhanced emissions in particular modes~\cite{Dicke1954}. This phenomenon, known as superradiance, highlights the coherent nature of spontaneous emission. On the other hand, it was clear since the first experiments~\cite{Skribanowitz1973,Gross1976} that several of its features could be understood through classical models~\cite{Gross1982}. Such classical analogues, however, cannot be applied to recent experiments observing the superradiant collective acceleration of emission with just a single excitation participating in the process~\cite{Mendes2013,deOliveira2014}. This single-photon superradiance is a direct manifestation of the wave-particle duality, with a single \textit{particle} being emitted faster due to the \textit{interference} of the probability amplitudes of emission by each atom. Such regime can be approximated by exciting an atomic sample with weak laser light~\cite{Roof2016,Araujo2016}, although the photon statistics in this scheme do not present quantum correlations.

Here we move further exploring superradiance with particular collective quantum states, and report its implementation in both the single- and two-excitations regime. We use the experimental scheme proposed in the Duan-Lukin-Cirac-Zoller (DLCZ) protocol for long distance quantum communication~\cite{Duan2001}, that resulted in a long line of works~\cite{Kuzmich2003,Eisaman2004,Balic2005,Matsukevich2005,Chou2005,Chou2007,Choi2010} exploring quantum correlations in the interaction of single photons with collective atomic memories. In our experiments, either one or two excitations are initially stored in the atomic memory. The readout process results in the superradiant emission of one or two photons, respectively, with properties that depend on the quantum state of the memory. Due to collective enhancement, the photon emission in the read process is highly directional, what permits an efficient detection by selecting the appropriate mode. Our main purpose is to observe the increase of the photons emission rate due to superradiance, together with the characterization of the Fock-state regimes with one or two photons being emitted by the memory. To do so, we measure the wavepackets of the single-photon and of the bi-photon emissions, evidencing superradiant acceleration in both cases, and perform a photon statistics analysis that indicates the presence of quantum correlations.

\begin{figure}[htb]
  \hspace{-0.4cm}\includegraphics[angle=270,width=9.0cm]{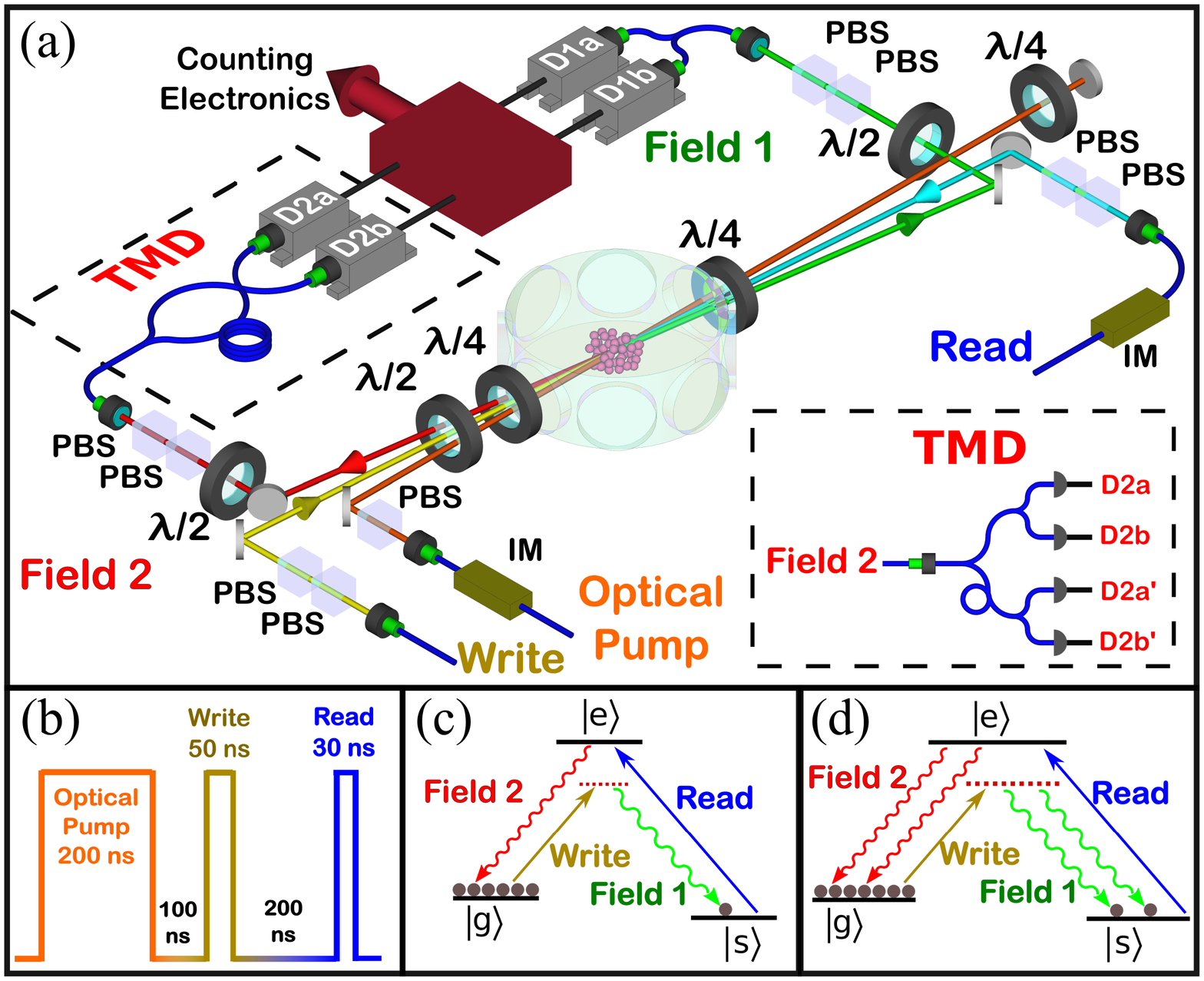}
  \vspace{-0.6cm}
  \caption{(colors online) (a) Experimental setup. PBS stands for Polarizing Beam Splitter; IM for in-fiber Intensity Modulator; TMD for Time-Multiplexing Detection; $\lambda/2$ and $\lambda/4$ for half- and quarter-wave plates, respectively. Inset shows the effective configuration of detectors of the TMD apparatus. (b) Pulse sequence for each measurement cycle. (c) and (d) provide the level scheme and fields for single- and two-photon superradiance, respectively.}\label{fig1}
\end{figure}

In the DLCZ scheme, an ensemble of  three-level atoms in $\Lambda$ configuration is initially prepared with all atoms in level $|g\rangle$ (Fig.~\ref{fig1}). A write beam induces the transition of atoms to level $|s\rangle$ through the emission of photons in a selected mode 1. The system state at this stage can be written as 
\begin{equation}
| \Psi_{a,1} \rangle = \sqrt{1-p} \sum_{n=0}^{\infty} p^{n/2} | n_a , n_1 \rangle \, , \label{e1}
\end{equation}
with storage of  $n$ excitations in a collective mode $a$ and $n$ photons in mode 1. The parameter $p$ indicates, for $p<<1$, the probability of having a single excitation both in the ensemble and in the light field. Using non-number-resolving detection with low efficiency (the usual case), a single detection in field 1 ideally projects the ensemble in the state
\begin{equation}
| \psi_1 \rangle \propto |1_a\rangle + p^{1/2} |2_a\rangle + p |3_a\rangle + \cdots \, . \label{e2}
\end{equation} 
On the other hand, two detections in field 1 would result in the state
\begin{equation}
| \psi_2 \rangle \propto |2_a\rangle + p^{1/2} |3_a\rangle + \cdots \, . \label{e3}
\end{equation} 
The Fock states $| 1_a\rangle$ and $|2_a\rangle$ are then obtained as limits of the states $|\psi_1\rangle$ and $|\psi_2\rangle$, respectively, when $p \rightarrow 0$. Differently from previous superradiant experiments, then, with this scheme we can investigate the phenomenon of superradiance with a controlled number of excited atoms in particular collective states.

The ensemble of cold rubidium 87 atoms in our experiment is obtained from a magneto-optical trap, turned off for 2~ms. After waiting 1~ms for the complete decay of the trap magnetic field, a sequence of 1000 sampling periods of 1~$\mu$s duration  follows. Residual DC magnetic fields are canceled following the method of Ref.~\onlinecite{deAlmeida2016}. The temperature of the atoms is below 1~mK so that their motion can be neglected during a sampling period. At each period an optical pumping field of 200ns duration (Figs.~\ref{fig1}a,\ref{fig1}b) prepares the atoms at the $|g\rangle = | 5S_{1/2}, F=2, m_F = -2 \rangle$ state. This beam is red-detuned $32\,$MHz from the $F=2 \rightarrow F^{\prime} = 3$ transition and has circular $\sigma^{-}$ polarization, being retro-reflected to reduce its mechanical action over the atoms. Pulse durations in the experiment are controlled by acousto-optic modulators and two 10 GHz in-fiber Mach-Zehnder Intensity Modulators (Fig.~\ref{fig1}a).

\vspace{-0.4cm}

\vspace{-0.3cm}
Once in state $|g\rangle$ the atoms are excited during $50\,$ns by a circular, $\sigma^{+}$ write pulse $22\,$MHz red-detuned from the $|g\rangle \rightarrow |e\rangle$ transition, with $|e\rangle = | 5P_{3/2}, F=2, m_F = -1 \rangle$. With small probability, $n$ atoms may be transferred to the state $|s\rangle = | 5S_{1/2}, F=1, m_F = 0 \rangle$ spontaneously emitting $n$ $\sigma^{-}$ photons in field 1 (Figs.~\ref{fig1}c,\ref{fig1}d). These are coupled to a single-mode Fiber Beam Splitter (FBS), leading to two detectors ($D_{1a}$, $D_{1b}$) for the projective measurements resulting in storage of $|\psi_1 \rangle$ or $|\psi_2\rangle$.

After a storage time of $200\,$ns, the atoms are excited by a strong, $30\,$ns read pulse resonant with $|s\rangle \rightarrow |e\rangle$. This pulse maps the stored collective state into the state of a second light mode, field 2, leaving the whole ensemble again in state $|g\rangle$. During the write process, atoms can also decay to other states, but these do not contribute to field 2~\cite{Mendes2013,Felinto2005}. Field 2 is then directed to the analysis by a Time-Multiplexing Detection (TMD) apparatus, consisting of a sequence of two FBS with a fiber loop delaying in $100\,$ns one of the arms in the middle. The outputs of the second FBS reach two detectors ($D_{2a}$,$D_{2b}$). This apparatus corresponds to a cascade of beam splitters leading to four detectors~\cite{Fitch2003}, as in the inset of Fig.~\ref{fig1}a, with $D_{2a}^{\prime}$,$D_{2b}^{\prime}$ representing the $100\,$ns delayed responses of $D_{2a}$,$D_{2b}$.

Field 1 is selected by an optical fiber in a gaussian mode with a 4$\sigma$ diameter of 150$\mu$m in the ensemble and forming an angle of about $2^{\rm o}$ with the direction of the write field, which has a 4$\sigma$ diameter of 420$\mu$m. The read and field-2 beams are mode matched and counter-propagating to the write and field-1 beams, respectively. This configuration results in single-mode superradiance with negligible propagation effects~\cite{Mendes2013,deOliveira2014}.

\begin{figure}[htb]
  \hspace{-0.25cm}\includegraphics[width=8.7cm]{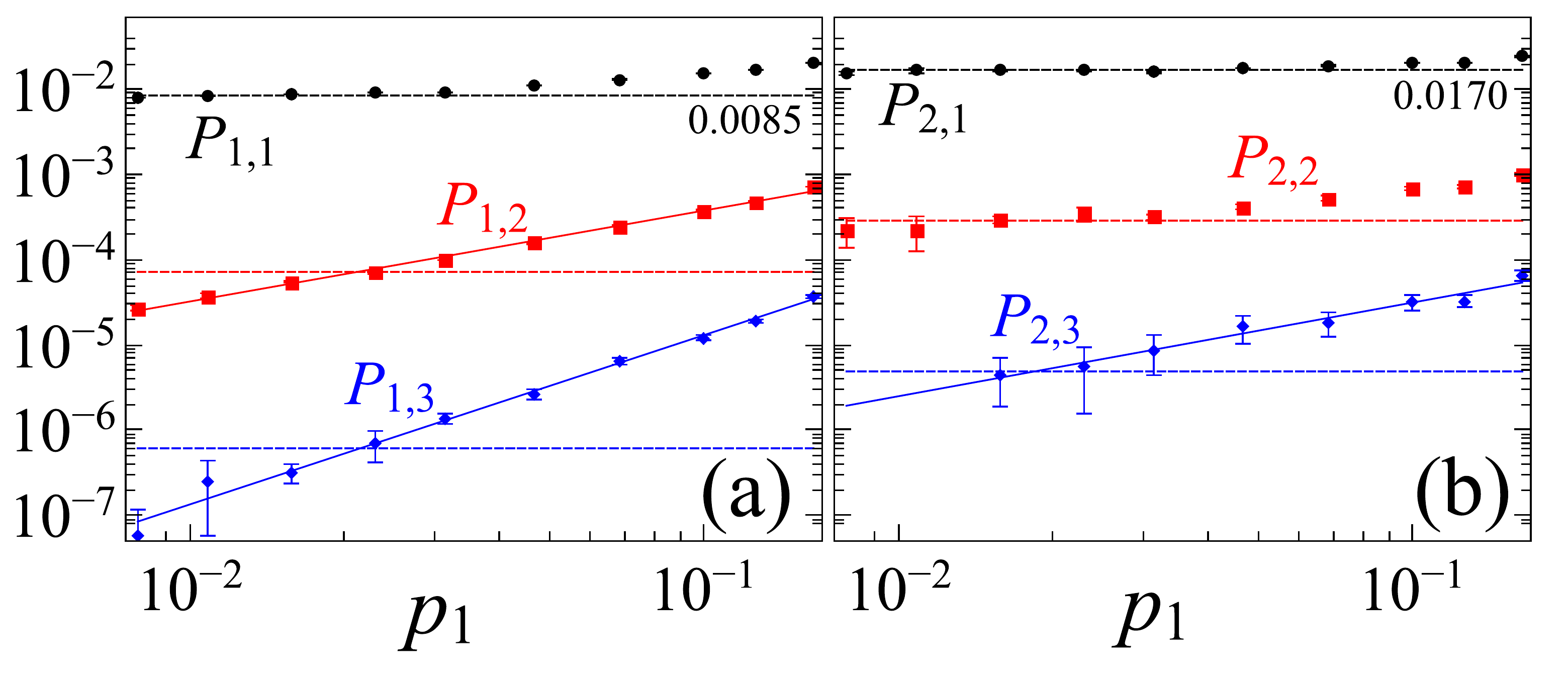}
  \vspace{-0.5cm}
  \caption{(colors online) Probabilities $P_{i,j}$ to detect $j$ photons in field 2 conditioned on the detection of $i$ photons in field 1 as a function of the probability $p_1$ to detect one photon in field 1, with $i=1$ (a) and $2$ (b). Circles, squares, and diamonds plot the probabilities of detecting one, two, and three photons in field 2. Solid lines are linear fits. Black dashed lines provide the values for the plateau of $P_{i,1}$ [0.0085 for (a) and 0.0170 for (b)]. Red and blue dashed lines gives the square and cube, respectively, of the black one, corresponding to the Poisson levels for two and three photons components.}\label{fig2}
\end{figure}

The photon-number analysis of field 2 conditioned on one (Fig.~\ref{fig1}c) or two (Fig.~\ref{fig1}d) detections in field 1 are presented in Figs.~\ref{fig2}a and~\ref{fig2}b, respectively, as a function of the probability $p_1$ for a detection in field 1 (ratio between number of detections in field 1 and number of trials). $P_{i,j}$ indicates the probability for $j$ detections in field 2 conditioned on $i$ detections in field 1. In this way, Fig.~\ref{fig2}a plots the values of $P_{1,j}$, related to $|\psi_1\rangle$, and Fig.~\ref{fig2}b the values of $P_{2,j}$, related to $|\psi_2\rangle$. The two panels were obtained from the same data set. Error bars come from the uncertainty in the accumulation of detection events, proportional to the square root of the number of detections.

\vspace*{-0.1cm}

\vspace*{-0.2cm}
To compare Fig.~\ref{fig2} to the predictions of Eq.~\eqref{e1}, note that $p_1 \approx \eta_1 p$, with $\eta_1$ the detection efficiency. As $p_1$ decreases, with decreasing write intensities, we observe two plateaus forming for $P_{1,1}$ and $P_{2,2}$, since those quantities should be roughly independent of $p$ in this limit (see Eqs.~\eqref{e2} and~\eqref{e3} for $|\psi_1\rangle$ and $|\psi_2\rangle$, respectively). For perfect detection, 100$\%$ efficiency and number resolving, we should not see a $P_{2,1}$. However, in our limit of low efficiency, the loss of a photon in the pair leads to a plateau on $P_{2,1}$ with twice ($\approx 0.017$) the value of $P_{1,1}$ ($\approx 0.0085$), since now two photons enter the TMD apparatus.

Fock states $|1_a \rangle$ and $|2_a \rangle$ are limits of $|\psi_1 \rangle$ and $|\psi_2\rangle$ when $p \rightarrow 0$. For finite $p$, there are always some higher order components. For instance, from Eq.~\eqref{e2} we expect the probabilities $P_{1,2}$ and $P_{1,3}$ to decrease proportionally to $p$ and $p^2$, respectively. From the log-log plot in Fig.~\ref{fig2}a, we obtain $P_{1,2} \propto p_1^{s_{12}}$ and $P_{1,3} \propto p_1^{s_{13}}$, with $s_{12} = 1.07 \pm 0.02$ and $s_{13} = 1.99 \pm 0.07$. From Eq.~\eqref{e3}, on the other hand, we expect $P_{2,3} \propto p$, obtaining $P_{2,3} \propto p_1^{s_{23}}$ with $s_{23} = 1.10 \pm 0.07$ from Fig.~\ref{fig2}b. Besides observing the predicted suppression of higher order components, it is also interesting to compare their values to the expectation for a coherent state with single-photon components consistent with the plateaus of Figs.~\ref{fig2}a,\ref{fig2}b (upper dashed lines). In both panels, the dashed lines in the middle and on the bottom give the square and the cube of the value for the upper line, the expected results for a coherent state. We measure then clear suppressions of $P_{1,2}$ and $P_{1,3}$ down to sub-poissonian levels. On the other hand, due to the low efficiency for detecting coincidences between five events (2 in field 1 and 3 in field 2), we could not measure $P_{2,3}$ in a clear sub-poissonian regime.

In order to directly address the superradiant aspects of the problem, we now focus on the wavepackets of the retrieved photons on field 2. Our guide is the theoretical expression for the single-photon wavepacket, given by the probability $p_c(t)$ to observe a single detection in field 2 at time $t$ after turning on the read pulse, conditioned on a single detection in field 1~\cite{deOliveira2014}:
\begin{equation}
\frac{p_c(t)}{P_c} = \alpha \, e^{-\chi\Gamma t/2} \sin^2\left( \frac{\Omega t}{2}\right) \Delta t \,, \label{pcPc}
\end{equation}
with $\alpha = \chi\Gamma\Omega_0^2/\Omega^2$, $\Omega = \sqrt{\Omega_0^2 - \chi^2\Gamma^2/4}$, $\Gamma /2\pi = 6.065$~MHz (natural linewidth of $|e\rangle$), $\Omega_0$ the reading Rabi frequency, and $\chi$ the superradiant enhancement for the $|e\rangle \rightarrow |g\rangle$ decay. $\Delta t$ is the detection window and $P_c$ the total detection probability, integrated in time. Equation~\eqref{pcPc} indicates the way to directly extract information on the superradiant acceleration from the wavepackets. In the high-read-power regime (high $\Omega_0$), the emission dynamics decouples into two well-defined parts~\cite{deOliveira2014}: a Rabi  oscillation between $|s\rangle$ and $|e\rangle$ with frequency close to $\Omega_0$ and an exponential decay from $|e\rangle$ to $|g\rangle$ with rate $\chi\Gamma$. Differently from a single atom decaying at a rate $\Gamma$, collective constructive interference in the ensemble may result in $\chi > 1$. Measuring the decay rate of the Rabi oscillations provides directly the value for $\chi \Gamma$. For the geometry of our ensemble, we calculate $\chi\approx 1+N/(\omega_0^2k^2)$ \cite{Mendes2013}, with $N$ the number of atoms in mode 1, $\omega_0$ the mode waist radius, and $k$ the photon wavenumber.  

\begin{figure}[htb]
  \hspace{0.0cm}\includegraphics[width=8.7cm]{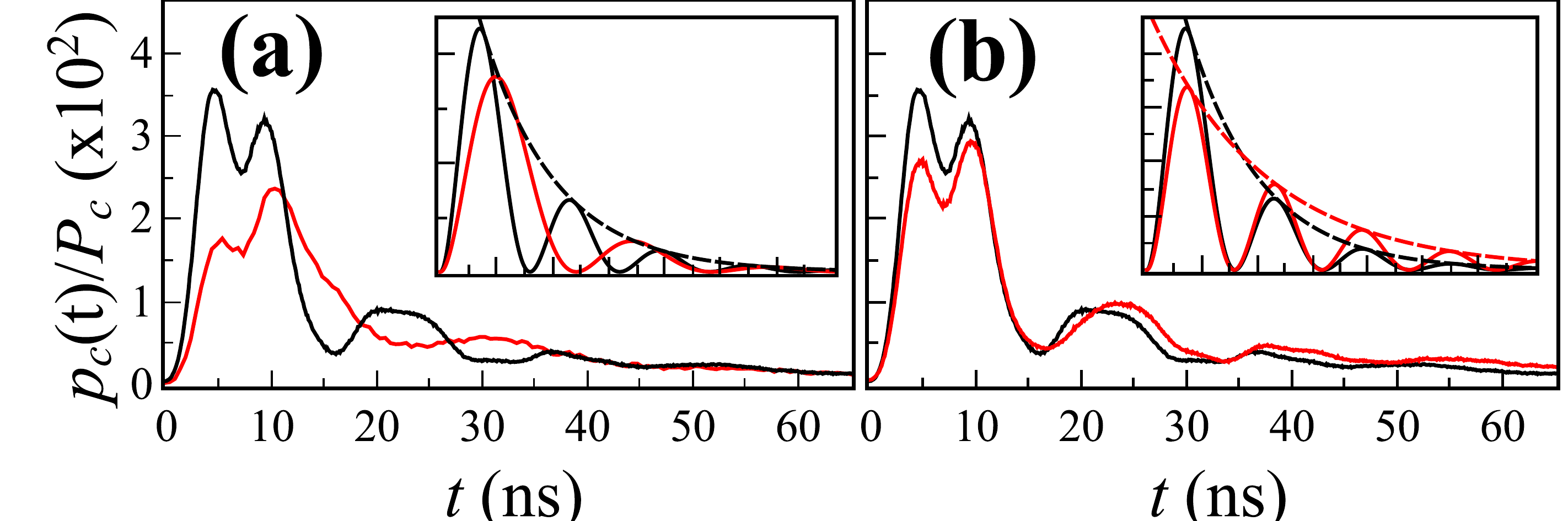}
  \vspace{-0.7cm}
  \caption{(colors online) {\bf Single-photon wavepackets.} Conditional probability $p_c(t)$ for a single detection in field 2 at time $t$, normalized by $P_c$. (a) For $OD_1$, wavepackets for $P_{R1} $ (black curve) and $P_{R2}$ (red curve). (b) For $P_{R1}$, wavepackets for $OD_1$ (black curve) and $OD_2$ (red curve). The insets plot the corresponding theoretical curves according to Eq.~\eqref{pcPc} along with the respective pure exponential decays (dashed lines).}\label{fig3}
\end{figure}

Two modifications had to be introduced on the setup of Fig.~\ref{fig1} at this stage. First, the read pulse duration was increased to 190~ns to fulfill the condition of constant read power assumed for Eq.~\eqref{pcPc}. Second, the TMD apparatus was substituted by a fiber beamsplitter, since the read duration is now longer than the fiber loop delay. For $|\psi_1\rangle$ we acquired three wavepackets, all plotted on Fig.~\ref{fig3}. The black curve on Figs.~\ref{fig3}a,~\ref{fig3}b is the wavepacket for our maximum Optical Depth ($OD_1 = 31.4\pm0.4$) and maximum read power ($P_{R1} = 3.95$~mW). Since $OD \propto N$, maximum $OD$ enhances the collective effects behind superradiance, while high read powers are crucial to induce Rabi oscillations. $OD$ was measured on the transition $|g\rangle \rightarrow |5P_{3/2},F=3,m_F=-3\rangle$, with $OD_1$ corresponding to $N \approx 1.9\times 10^6$~\cite{deOliveira2014}. Other important parameters for this curve are $\Delta t = 0.5$~ns, $P_c = 6.3\%$, $p_1 = 0.015$, and $g_2 = 0.405\pm 0.004 \approx P_{1,2}/P_{1,1}^2$, indicating a two-photon component significantly below the poissonian level of 1. The insets provide the corresponding theoretical predictions from Eq.~\eqref{pcPc}. For the black curve on the insets we assumed $\Omega_0 = 0.4\times 10^9$rad/s (to match the observed Rabi oscillation) and $\chi = 4.0$ (to match the observed decay of the oscillations). These theoretical values would correspond to $P_R = 2.1$~mW and $N \approx 1.1\times 10^6$~\cite{deOliveira2014}, lying within a factor of two of our estimation for these experimental parameters.

\vspace*{-0.1cm}

The red curve on Fig.~\ref{fig3}a represents the photon wavepacket with $OD_1$ and the read power reduced to $P_{R2} = 1.76$~mW. For the red curve on the inset of panel (a) we modify the Rabi frequency to $\Omega_0 =0.27\times 10^9$rad/s, consistent with the read-power relation $\sqrt{P_{R2}/P_{R1}} \approx 0.67$ between red and black curves. It can be seen that the modification of the read power changes the frequency of the Rabi oscillations, but not the exponential decay rate. The red curve on Fig.~\ref{fig3}b represents the photon wavepacket with $P_{R1}$ and the optical depth reduced to $OD_2 = 15.9\pm0.5$. The optical depth was decreased by reducing the trapping laser power~\cite{deOliveira2014}. On the inset of panel (b) the red curve is for $\chi = 2.52 = 1+ (4.0-1)(OD_2/OD_1)$, since $\chi -1$ and $OD$ are both proportional to $N$~\cite{Mendes2013,deOliveira2014}. It can be seen that the modification of the number of atoms changes the superradiant decay rate, but not the frequency of the Rabi oscillations.

The single-photon wavepackets on Fig.~\ref{fig3} follow Eq.~\eqref{pcPc} with a few remarks. For start, the first minimum of the experimental curves, at $t = 7$~ns, has no relation to the underlining dynamics we are investigating, coming from a small ringing on the beginning of the read pulse that we were not able to fully eliminate. Note that its temporal position does not vary with read power or optical depth. Second, the single-photon wavepacket reaches a small plateau for long times. This comes from a larger noise level due to removal of a frequency filter in field 1~\cite{deOliveira2014}, resulting in significant increase in photon-pair generation rate, up to 40Hz. Finally, the number of atoms was changed by a relatively small amount between the black and red curves in Fig.~\ref{fig3}b, to avoid decreasing the rate of four-photons detections. Both compromises to improve the count rates were crucial for  the two-photon wavepackets measurements.  

\vspace*{-0.0cm}
\begin{figure}[htb]
  \hspace{0.0cm}\includegraphics[width=8.5cm]{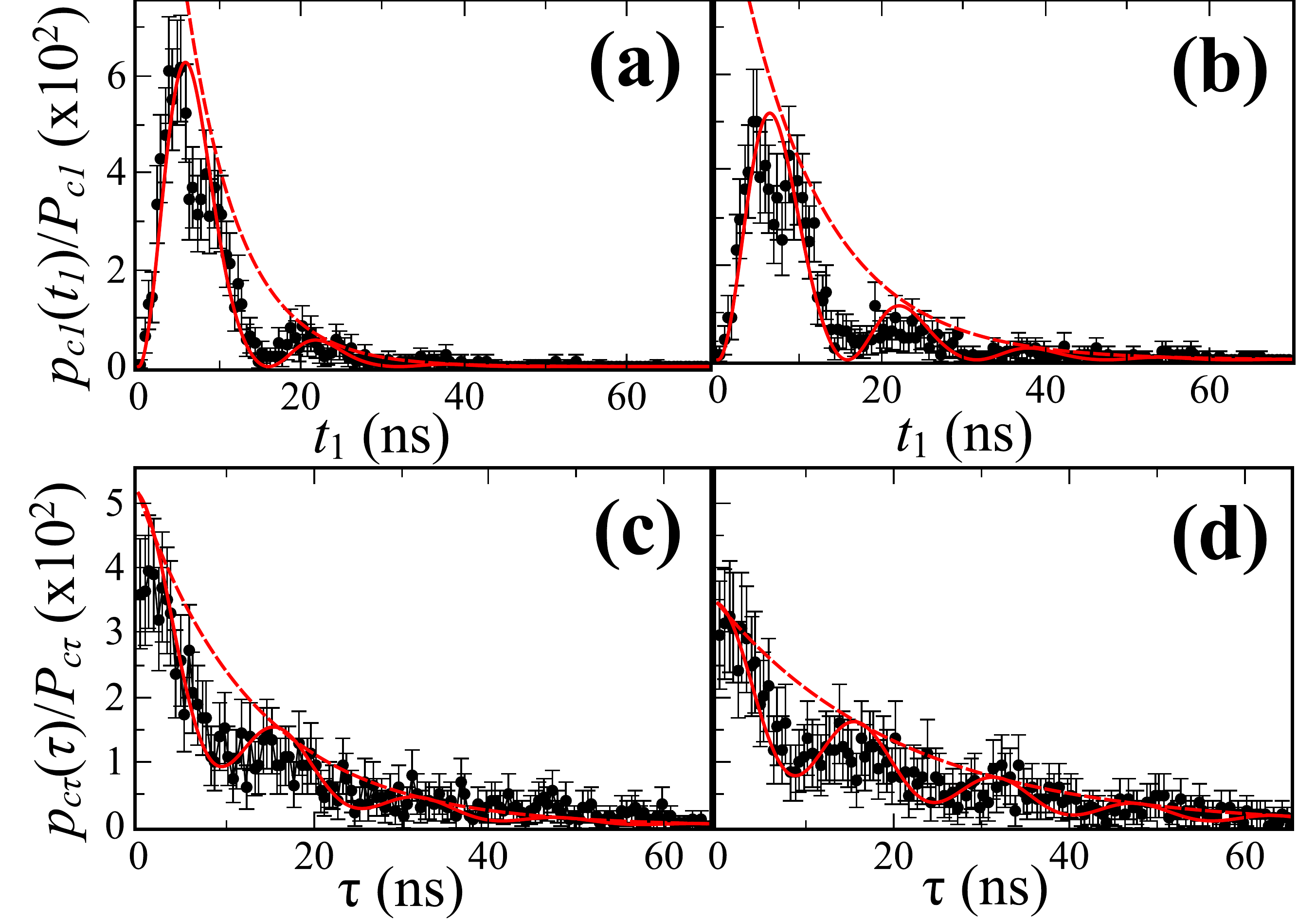}
  \vspace{-0.3cm}
  \caption{(colors online) {\bf Biphoton wavepackets.} Panels (a) and (b): Probability $p_{c1}(t_1)$ to detect the first field-2 photon at $t_1$, normalized by $P_{c1}$. Panels (c) and (d): Probability $p_{c\tau}(\tau)$ to detect the second photon at time $\tau$ after the first detection, normalized by $P_{c\tau}$. Data on panels (a) and (c) [(b) and (d)] resulted from the same measurements as for the experimental black [red] curve on Fig.~\ref{fig3}b. Solid lines provide the theory from Eqs.~\eqref{e5} and \eqref{e6}, for the same parameters as the inset in Fig.~\ref{fig3}b. Dashed lines plot the respective pure exponential decays.}\label{fig4}
\end{figure}

\vspace*{-0.1cm}
Our measurements for the superradiant two-photon wavepackets, in which we observe two detections in field 2 conditioned on two detections in field 1, are shown in Fig.~\ref{fig4} for the two optical depths of Fig.~\ref{fig3}b. Figures~\ref{fig4}a,\ref{fig4}c were obtained for $OD_1$, and Figs.~\ref{fig4}b,\ref{fig4}d for $OD_2$. Since there are two detections in field 2, the wavepacket information was divided in two parts. In panels (a),(b) we plot the probability $p_{c1}(t_1)$ of detecting the first photon of the pair in field 2 at a time $t_1$ after turning on the read field. In panels (c),(d) we plot the conditional probability $p_{c\tau}(\tau)$ of detecting the second photon of the pair at a time $\tau$ after the first one. Our largest rate of four-photon generation, for $OD_1$, was 14~mHz. 

Neglecting the reabsorption of photons by the atomic ensemble, a theoretical analysis of the reading process starting with a Fock state $|2_a\rangle$ in the atomic ensemble leads to a simple result for the two-photon wavepacket~\cite{Davi2017}: $p_{cc}(t_1,t_2) = p_c(t_1)p_c(t_2)$. This approximation is justified in our system~\cite{deOliveira2014} since the read process occurs in a condition of Electromagnetically Induced Transparency (EIT)~\cite{Boller1991}, in which the read pulse induces a transparency in the medium for the outgoing photons.

From Eq.~\eqref{pcPc} and the simple expression relating it to the two-photon wavepacket, we derive the theoretical functions~\cite{Davi2017} to directly compare to the experimental results of Fig.~\ref{fig4}. By having $t_2 = t_1+\tau > t_1$, we can integrate over $\tau$ to obtain the normalized probability to detect in $t_1$ the first photon of the pair in field 2:
\begin{align}
\frac{p_{c1}(t_1)}{P_{c1}} &= a_1e^{-\chi\Gamma t_1}\sin^2\left( \frac{\Omega t_1}{2}\right) \times\nonumber \\ 
&\times \left[ 1 + b_1 \sin\left(\Omega t_1\right) - c_1 \cos\left(\Omega t_1\right)\right]\, , \label{e5}
\end{align}
where $a_1=2\,\alpha\Delta t_1$, $b_1= \chi\Gamma\Omega/2\Omega_0^2$, $c_1 = (\chi\Gamma)^2/4\Omega_0^2$, and $P_{c1}$ is a constant equal to the integral of the curve for $p_{c1}(t_1)$. On the other hand, if the integration run over $t_1$, we obtain the probability to detect the second photon in field 2 at a time $\tau$ after the first one was detected:
\begin{align}
&\frac{p_{c\tau}(\tau)}{P_{c\tau}} = a_{\tau}e^{-\chi\Gamma \tau/2} [1+b_{\tau}\cos\left(\Omega \tau\right) +c_{\tau}\sin\left(\Omega \tau\right) ],\label{e6}
\end{align}
with $a_{\tau}=\alpha\Delta \tau\Omega_0^2/(2\Omega^2+2\chi^2\Gamma^2)$, $b_{\tau}=(6\Omega^2-4\Omega_0^2)/4\Omega_0^2$, $c_{\tau} = 3\Omega\chi\Gamma/4\Omega_0^2$, and $P_{c\tau}$ another normalization constant to keep the integral of the curve equal to one. Equation~\eqref{e5} [\eqref{e6}] is plotted as the red curves on Figs.~\ref{fig4}a and~\ref{fig4}b [\ref{fig4}c and~\ref{fig4}d], for the same parameters, respectively, as the black and red curves of Fig.~\ref{fig3}b. The results of Eqs.~\eqref{e5} and~\eqref{e6}  capture the essential aspects of the two-photon wavepackets, with the decay of $p_{c1}(t_1)$ with twice the rate of $p_{c\tau}(\tau)$ and Rabi oscillations in both curves. This clearly demonstrates the superradiant emission of the biphoton, with the proper enhanced decay rates, and largely validates our hypothesis of independence in the emission of the two photons. These results are consistent with Dicke's theory for superradiance \cite{Dicke1954}, which also neglects interactions between the outgoing photons.

In conclusion, we have shown single-photon and two-photon superradiance in the reading process of an atomic memory. The photon statistics analysis confirmed that the emitted light was close to Fock states, and the temporal dynamics of the photons emissions confirmed that they were in superradiant regimes. There are still plenty of room to improve the system both in terms of four-photon generation rate and number of atoms in the ensemble. Larger generation rates may lead to purer single- and two-excitation states, but also to investigations of collective states with three or more excitations. On the other hand, larger numbers of atoms, through larger optical depths of the atomic ensemble, may lead to different superradiant regimes, possibly presenting some interaction between the extracted photons. As a whole, such developments point to the feasibility of a new approach to generate and control larger and purer Fock states connected to long-lived atomic memories, useful for quantum metrology~\cite{Holland1993} and helping to lift the usually assumed restriction to single-photon sources as a possible resource in the designing of new quantum information protocols~\cite{vanLoock2011}.

\begin{acknowledgments}

This work was supported by CNPq, CAPES, FAPEMIG and FACEPE (Brazilian agencies), through the programs PRONEX and INCT-IQ (Instituto Nacional de Ci\^encia e Tecnologia de Informa{\c c}\~ao Qu\^antica).

L. O.-G. and L. F. M.-M. contributed equally to this work.

\end{acknowledgments}


\begin{thebibliography}{16}

\bibitem{Dicke1954}
R. H. Dicke, Phys. Rev. {\bf 93}, 99 (1954).

\bibitem{Skribanowitz1973}
N. Skribanowitz, I. P. Herman, J. C. MacGillivray, and M. S. Feld, Phys. Rev. Lett. {\bf 30}, 309 (1973).

\bibitem{Gross1976}
M. Gross, C. Fabre, P. Pillet, and S. Haroche, Phys. Rev. Lett. {\bf 36}, 1035 (1976).

\bibitem{Gross1982}
M. Gross and S. Haroche, Phys. Rep. {\bf 93}, 301 (1982).

\bibitem{Mendes2013}
M. S. Mendes, P. L. Saldanha, J. W. R. Tabosa, and D. Felinto, New J. Phys. {\bf 15}, 075030 (2013).

\bibitem{deOliveira2014}
R. A. de Oliveira, M. S. Mendes, W. S. Martins, P. L. Saldanha, J. W. R. Tabosa, and D. Felinto, Phys. Rev. A {\bf 90}, 023848 (2014).

\bibitem{Roof2016}
S. J. Roof, K. J. Kemp, M. D. Havey, and I. M. Sokolov, Phys. Rev. Lett. {\bf 117}, 073003 (2016).

\bibitem{Araujo2016}
M. O. Araujo, Ivor Kresic, Robin Kaiser, and William Guerin, Phys. Rev. Lett. {\bf 117}, 073002 (2016).

\bibitem{Duan2001}
L.-M. Duan, M. D. Lukin, J. I. Cirac, and P. Zoller, Nature {\bf 414}, 413 (2001).

\bibitem{Kuzmich2003}
A. Kuzmich, W. P. Bowen, A. D. Boozer, A. Boca, C. W. Chou, L.-M. Duan, and H. J. Kimble, Nature {\bf 423}, 731 (2003).

\bibitem{Eisaman2004}
M. D. Eisaman, L. Childress, A. Andr{\'e}, F. Massou, A. S. Zibrov, and M. D. Lukin, Phys. Rev. Lett. {\bf 93}, 233602 (2004).

\bibitem{Balic2005}
V. Balic, D. A. Braje, P. Kolchin, G. Y. Yin, and S. E. Harris, Phys. Rev. Lett. {\bf 94}, 183601 (2005).

\bibitem{Matsukevich2005}
D. N. Matsukevich, T. Chaneli{\`e}re, M. Bhattacharya, S.-Y. Lan, S. D. Jenkins, T. A. B. Kennedy, and A. Kuzmich, Phys. Rev. Lett. {\bf 95}, 040405 (2005).

\bibitem{Chou2005}
C. W. Chou, H. de Riedmatten, D. Felinto, S. V. Polyakov, S. J. van Enk, and H. J. Kimble, Nature {\bf438}, 828 (2005).

\bibitem{Chou2007}
C.~W. Chou, J. Laurat, H. Deng, K. S. Choi, H. de Riedmatten, D. Felinto, and H. J. Kimble, Science {\bf 316}, 1316 (2007).

\bibitem{Choi2010}
K. S. Choi, A. Goban, S. B. Papp, S. J. van Enk, and H. J. Kimble, Nature {\bf 468}, 412 (2010).
\bibitem{deAlmeida2016}
A. J. F. de Almeida, M.-A. Maynard, C. Banerjee, D. Felinto, F. Goldfarb, and J. W. R. Tabosa, Phys. Rev. A {\bf 94}, 063834 (2016).

\bibitem{Felinto2005}
D. Felinto, C. W. Chou, H. de Riedmatten, S. V. Polyakov, and H. J. Kimble, Phys. Rev. A {\bf 72}, 053809 (2005).

\bibitem{Fitch2003}
M. J. Fitch, B. C. Jacobs, T. B. Pittman, and J. D. Franson, Phys. Rev. A {\bf 68}, 043814 (2003).

\bibitem{Davi2017}
D. F. Barros, L. G. O. Gutierrez, L. F. M. Martinez, J. E. O. Morales, R. S. N. Moreira, N. D. Alves, A. F. G. Tieco, D. Felinto and P. L. Saldanha, in preparation. 

\bibitem{Boller1991}
K.-J. Boller, A. Imamoglu, and S. E. Harris, Phys. Rev. Lett. {\bf 66}, 2593 (1991).

\bibitem{Holland1993}
M. J. Holland and K. Burnett, Phys. Rev. Lett. {\bf 71}, 1355 (1993).

\bibitem{vanLoock2011}
P. van Loock, Laser Photonics Rev. {\bf 5}, 167 (2011).

\end{thebibliography}
\end{document}